\documentclass[12pt]{article}
\usepackage{subeqn}
\usepackage{epsfig}
\newcommand{\be}{\begin{equation}}
\newcommand{\ee}{\end{equation}}

\newcommand{\no}{\noindent}
\newcommand{\ce}{\begin{center}}
\newcommand{\nc}{\end{center}}

\makeatletter
\@addtoreset{equation}{section}
\makeatother

\def\sqr#1#2{{\vcenter{\vbox{\hrule height.#2pt
 \hbox{\vrule width.#2pt height#1pt \kern#1pt
 \vrule width.#2pt} \hrule height.#2pt}}}}

\def\operp{\hbox{${\kern+.25em{\bigcirc}
\kern-.85em\bot\kern+.85em\kern-.25em}$}}

\def\lsim{\;\raise0.3ex\hbox{$<$\kern-0.75em\raise-1.1ex\hbox{$\sim$}}\;}
\def\gsim{\;\raise0.3ex\hbox{$>$\kern-0.75em\raise-1.1ex\hbox{$\sim$}}\;}
\def\no{\noindent}

\def\ce{\centerline}
\def\ve{\vfill\eject}
\def\rdots{\mathinner{\mkern1mu\raise1pt\vbox{\kern7pt\hbox{.}}\mkern2mu
 \raise4pt\hbox{.}\mkern2mu\raise7pt\hbox{.}\mkern1mu}}

\def\e e{$e^+ e^-$ }

\begin{document}

\ce{\bf FLAVOR STATES OF THE KNOT MODEL}
\vskip.3cm

\ce{\it Robert J. Finkelstein}
\vskip.3cm

\ce{Department of Physics and Astronomy}
\ce{University of California, Los Angeles, CA 90095-1547}

\vskip1.0cm

\no {\bf Abstract.}  We discuss flavor states of the knot model and
their relation to the CKM and the PMNS matrices.  These states are
eigenstates of absorption-emission operators and are analogous to the
coherent states of the Maxwell field.  The underlying model has been
proposed as a possible substructure of the standard model.  We include a knot parametrization of the CKM matrix.

\ve

\section{Introduction}

To describe the weak decays of the strange particles Cabibbo
introduced a $2\times 2$ mixing matrix that was later extended by
Kobayashi and Maskawa to the $3\times 3$ matrix labelled by the
three flavor states of the up and down quarks.  Following the
suggestion of Pontecorvo the neutrino oscillations may similarly
be described by the PMNS matrix that expresses the three flavor
states as mixtures of the three mass states of the three leptonic 
neutrinos.$^1$

Flavor states are eigenstates of the absorption-emission operators that in
turn depend on the dynamics of the model.  Since we are here
interested in the flavor states of the knot model, we shall first
summarize the kinematical structure of this model as determined by
the symmetry algebra, $SL_q(2)$, in preparation for introducing the
dynamical assumptions, that are also subject to $SL_q(2)$, and that
determine the flavor states.

\vskip.5cm

\section{Quantum Trefoils$^2$}

We require that a quantum knot be described by one member of an
irreducible representation of the knot algebra $(SL_q(2))$, which is
here denoted by $D^j_{mm^\prime}$.  It is also required that there be
a correspondence between $D^j_{mm^\prime}$, and a classical knot.
Both requirements are met by allowing only those elements
$D^j_{mm^\prime}$, to represent quantum knots for which
\be
(j,m,m^\prime) =  \frac{1}{2} (N,w,r+1)
\ee
where $(N,w,r)$ are the number of crossings, the writhe, and the
rotation that describe the projection of a 3-dimensional classical
knot onto a 2-dimensional plane.  The simplest classical knot is the
trefoil having the 2d projection described by
\be
(N,w,r) = (3,\pm 3,\pm 2)
\ee
By (2.1) the corresponding four quantum trefoils are represented by
\be
D^{3/2}_{\frac{3}{2}\frac{3}{2}},~ D^{3/2}_{-\frac{3}{2}\frac{3}{2}},~
D^{3/2}_{\frac{3}{2}-\frac{1}{2}},~ D^{3/2}_{-\frac{3}{2}-\frac{1}{2}}
\ee
There are four quantum trefoils but only two of the four corresponding
classical trefoils can be topologically distinguished.  Note also
that $2m$ and $2m^\prime$, belonging to the same representation, are
of same parity while $w$ and $r$, describing the projection of a
classical knot, are required to be of opposite parity.

\vskip.5cm

\section{Irreducible Representations of the Knot Algebra${}^2$}

The  $2j+1$ dimensional representation of  $SLq(2)$  may be written as
follows:

\be
D^j_{mm^\prime}(a,b,c,d) = \sum_{\scriptstyle 0\leq s\leq n_+
\atop\scriptstyle 0\leq t\leq n_-}
{\cal A}^{j}_{mm^\prime}(q,s,t)
\delta(s+t,n^\prime_+)
a^s b^{(n_{+}-s)}c^{t} d^{(n_{-}-t)}
\ee
where
\be
n_\pm = j \pm m  
\ee
\be
n_\pm ^\prime = j \pm m'  
\ee
and the arguments  $(a,b,c,d)$  satisfy the knot algebra:${}^2$
\[
\begin{array}{llllr}
ab= qba & bd=qdb & bc = cb & ad-qbc=1 \\
ac =qca & cd=qdc & & da-q_{1}cb=1 \hspace{1.25in}\mbox{(A)}
\end{array}
\]

\noindent
where  $q_1 = q^{-1}$.

The  ${\cal A} _{mm'}^j $  are  $q$-deformations of the Wigner coefficients
that appear in irreducible representations of  $SU(2)$.

The knot algebra  (A)  and hence  $D_{mm'}^j (a,b,c,d)$  are defined
only up to the gauge transformation
\be
U_a (1): \begin{array}{l}
a^\prime =e^{i\varphi_a} a \\
d^\prime = e^{-i\varphi _a}d 
  \end{array}     \;\;\;\;\;\;\;\;\;\;\;\;\;\;
U_b (1): \begin{array}{l}
b^\prime =e^{i\varphi_b}b \\
c^\prime = e^{-i\varphi _b} c 
  \end{array}      
\ee

Eqns. (3.4) leave the algebra  $(A)$  invariant and induce on the elements
of every representation the following  $U_a (1) \times U_b(1)$  gauge
transformation
\be
D_{mm'}^{j}(a^{\prime}, b^{\prime}, c^{\prime}, d^{\prime})= e^{i(\varphi_a + \varphi _b)m} \;  
e^{i(\varphi _a - \varphi _b) m^{\prime}} \; D _{mm'}^{j}(a,b,c,d) 
\ee

\section{Field Theory and Charges of Quantum Trefoils$^2$}

One may construct a field theory of the quantum knots  
by attaching  $D_{mm'}^j$  to a standard 
field operator  $\psi (x)$  as follows:
\be
\Psi _{mm'}^j = \psi (x)  \ D _{mm'}^j 
\ee
By (3.5),  the field operator  $\Psi _{mm'}^j $ also transforms under
the gauge transformations  $U_a (1) \times U_b (1) $.
If the attachment (4.1) is made consistently for both fermionic and
bosonic fields one may construct a modified standard action that is
invariant under $U_a\times U_b$, as is shown in
Ref. 3.   This invariance of the field action is a
physical requirement
since the relabelling of the algebra described by  (3.4)  cannot
affect the physics.  Then in view of this invariance there will be 
by Noether's theorem one
conserved charge associated with  $U_a (1)$ and a second conserved charge
associated with  $U_b (1)$.  Then by (3.5) and (2.1) these charges may be defined by
\begin{eqnarray}
Q(w) \equiv  -k_w m &=& -k_w \ \frac{w}{2}  \\
Q(r) \equiv - k_r m' &=& - k_r \ \frac{r+1}{2}   
\end{eqnarray}
and may be referred to as the writhe and rotation charges.  
Here  $k_w$
and  $k_r$  are undetermined constants with the dimensions of an
electric charge.  In terms of  $Q(w)$  and  $Q(r)$,  the  $U_a(1) \times U_b(1)$
transformations on  $\Psi _{mm'}^j$  become
\be
\Psi _{mm'}^{j^\prime } = e ^{\frac{-i}{k_w} Q(w) \varphi (w) }
e^{\frac {-i}{k_r} Q(r) \varphi (r)} \ \Psi _{mm'}^j 
\ee
where  $\varphi (w) = \varphi _a + \varphi _b $  and  $\varphi (r) =
\varphi _a - \varphi _b$  by (3.5).

Since we expect that the most elementary particles, 
the elementary fermions, are quantum trefoils
in any natural knot model,   
we next make a direct comparison
between the  $Q(w)$  and  $Q(r)$  charges of the 
four quantum trefoils and the
charge and hypercharge of the four fermion families of the standard
theory where the members of each family are denoted by  
$(f_1, f_2, f_3 )$  in Table 4.1.${}^2$ 
The knot entries in the table are determined by (2.1), (4.2),
and (4.3).

\begin{table}[h]
\begin{center}
{\bf{Table 4.1}}
\end{center}
\begin{center}
\begin{tabular}{ccccc|ccccc}
\multicolumn{5}{c}{\underline{Standard Representation}} &
\multicolumn{5}{c}{\underline{Trefoil Representation}} \\
\underline{$(f_1,f_2,f_3)$} & \underline{$t$} &
\underline{$t_3$} &
\underline{$t_0$} & \underline{$Q_e$} & \underline{$(w,r)$} &
\underline{$D^{N/2}_{\frac{w}{2}\frac{r+1}{2}}$} &
\underline{$Q_w$} & \underline{$Q_r$} & \underline{$Q_w+Q_r$} \\
$(e,\mu,\tau)_L$ & $\frac{1}{2}$ & $-\frac{1}{2}$ & $-\frac{1}{2}$
& $-e$ & (3,2) & $D^{3/2}_{\frac{3}{2}\frac{3}{2}}$ &
$-k\left(\frac{3}{2}\right)$ & $-k\left(\frac{3}{2}\right)$ &
$-3k$ \\
$(\nu_e,\nu_\mu,\nu_\tau)_L$ & $\frac{1}{2}$ & $\frac{1}{2}$ &
$-\frac{1}{2}$ & 0 & (-3,2) & $D^{3/2}_{-\frac{3}{2}
\frac{3}{2}}$ & $-k\left(-\frac{3}{2}\right)$ &
$-k\left(\frac{3}{2}\right)$ & 0 \\
$(d,s,b)_L$ & $\frac{1}{2}$ & $-\frac{1}{2}$ & $\frac{1}{6}$ &
$-\frac{1}{3}e$ & (3,-2) & $D^{3/2}_{\frac{3}{2}-\frac
{1}{2}}$ & $-k\left(\frac{3}{2}\right)$ & $-k\left(-\frac{1}{2}
\right)$ & $-k$ \\
$(u,c,t)_L$ & $\frac{1}{2}$ & $\frac{1}{2}$ & $\frac{1}{6}$ &
$\frac{2}{3}e$ & (-3,-2) & $D^{3/2}_{-\frac{3}{2}
-\frac{1}{2}}$ & $-k\left(-\frac{3}{2}\right)$ &
$-k\left(-\frac{1}{2}\right)$ & $2k$ \\
\end{tabular}
\end{center}
\end{table}

In Table (4.1)  we have assumed a single value of  $k$:
\be
k_r = k_w = k 
\ee
which is also the same for all trefoils.  If we set  $k=e/3$, we find
that the charges of the four fermion families are related to the charges of the four quantum trefoils
as follows:
\begin{eqnarray}
Q_w = e t_3 \\ 
Q_r = e t_0 \\ 
Q_w + Q_r = Q_e 
\end{eqnarray}
in agreement with the standard model where there is the independent
relation for the electric charge
\be
Q_e = e(t_3+ t_0) 
\ee

If one aligns the trefoils and the fermion families in any order
different from that in Table 4.1, one needs more than a single value of  $k$ to relate $(t_3,t_0)$ to $(Q_w,Q_r)$.  
It is important that we choose  $k_r = k_w$  and that we also
choose a single value of  $k$  for the four quantum trefoils.  Note that
it is also not possible to exchange  $t_3 $  and  $t_0$  in (4.6) and
(4.7).  Therefore the correspondence between the four fermion families
and the four trefoils, is
empirically fixed and unique.  The value of $k$ as $e/3$ follows from
the identification of the total charge of the trefoil,
$Q_w+Q_r$, with $Q_e$.  One may also read directly from the table
\be
(j,m,m^\prime) = 3(t,-t_3,-t_0)
\ee
and enter the information conveyed by (4.10) into (4.1) as follows:
\be
\Psi^{3/2}(t_3,t_0,n) = \psi(t_3,t_0,n) D^{3/2}_{-3t_0-3t_0}|n\rangle
\ee
where $\psi(t_3,t_0,n)$ is the quantum field of the standard model
that represents the fermion with electroweak $SU(2)\times U(1)$ quantum numbers $(t_3,t_0)$.  Here $|n\rangle$ lies in the state space
defined by the knot algebra where $n=0,1,2$ labels the generation,
e.g. $(e,\mu,\tau)$.  Then $D^{3/2}_{-3t_3-3t_0}|n\rangle$ may be
regarded as an ``internal state function" reminiscent of a classical
knot and providing substructure to the elementary quantum fields of
the standard model.

We shall now propose that the non-trivial correspondence embodied in
Table (4.1) and expressed by (4.15) for the elementary fermions holds
more generally in the following form
\be
\Psi _{t_3 t_0}^t (n)=  \psi (t,t_3, t_0, n) D_{-3t_3-3t_0}^{3t} |n> 
\ee
i.e., we assume that  $(t,t_3,t_0)$  are related to  $(j,m,m')$  just as
in the special case $t=\frac12$:
$$
3t = j  \eqno(4.13j)
$$
$$
3t_3 = -m  \eqno(4.13m)
$$
$$
3t_0 = -m' \eqno(4.13m^\prime)
$$

\noindent
In other words we assume that there is an underlying  $SLq(2)$ symmetry
of the elementary particles that may be expressed through the internal
state functions $  D_{mm'}^j \left. | n\right> $.  For  $j\geq 1$  not all states  
$(m,m')$  of  $D_{mm'} ^j $ are filled.  The occupied states are
labelled by $ D_{-3t_3-3t_0}^{3t}$  according to  (4.13)  and are
determined by the intersection of the electroweak  $SU(2)\times U(1)$ and
the  $SL_q(2)$ symmetries.
For example, the
analogue of Table 4.1 for the elementary fermions is Table 4.2 for the
elementary bosons of the Weinberg-Salam model.  The $|n\rangle$ 
in (4.12) are
intended to represent the possible states of excitation of the 
quantum knot.

\begin{table}[h]
\begin{center}
\bf{Table 4.2}
\end{center}
\begin{center}
\begin{tabular}{ccccc}
    &\underline{$t$} & \underline{$t_3$} & \underline{$t_0$} & 
                \underline{$D_{-3t_3 -3k_0}^{3t}$}\\
$W^+$  & $1$   &  $1$ & $0$  & $D_{-30}^3$ \\
$W^-$  & $1$   & $-1$  & $0$  & $D_{30}^3$ \\
$W^3$  & $1$   & $ 0$ & $0$ & $D_{00}^3$ \\
$W^0$  & $0$   & $ 0$  & $ 0$ &  $D_{00}^0$
\end{tabular}
\end{center}
\end{table}

We adopt the following rule:
If a particle is labelled in the standard model by electroweak quantum
numbers  $(t,t_3,t_0)$  then attach to the quantum field operator of
that particle the factor  $D_{-3t_3 -3t_0}^{3t} (a,b,c,d)$.  
This factor
is to be understood as an element of the $j=3t$  representation of the  $SL_q(2)$  algebra and may be interpreted as the replacement of the point
particle of the standard model by a solitonic structure described solely
by this factor.  The extension of (4.11) to (4.12) expresses the
conservation of $t_3$ and $t_0$ everywhere in the modified 
model as a 
joint consequence of the $U_a\times U_b$ and the
$SU(2)\times U(1)$ invariance.

\vskip.5cm

\section{The Electroweak Interactions$^4$}

In the  $SLq(2)$ model the solitonic fermions interact by the emission
and absorption of solitonic bosons.  Denote the generic fermion-boson
interaction by
\be
\bar{{\cal  F}}'' \;  {\cal B}' \;  {\cal F}   
\ee
where
\be
{\cal F} = \mbox{{\bf F}}(p,s,t_3, t_0) \left( D_{-3t_3 - 3t_0}^{3/2} \right)  |n>
\ee
\be
\bar{{\cal F}}''  = <n''| \left( \bar{D}_{-3t_3 -3t_0} ^{3/2} \right) '' \;
\bar{\mbox{{\bf F}}}''(p,s,t_3, t_0 )
\ee
\be
{\cal B}' = \mbox{\bf B}'(p,s,t_3, t_0 ) \left( D_{-3t_3-3t_0} ^{3t} \right)'
\ee
and the pair $(p,s)$ refer to momentum and spin.  
Then (5.1) becomes
\be
(\bar{\mbox{{\bf F}}}'' \mbox{{\bf B}}' \mbox{{\bf F}}) 
<n''| \bar{D}_{ -3t''_{3} -3t''_{0}}^{3/2} \;
D_{-3t'_3- 3t'_0}^{3t} \; D_{-3t_3-3t_0}^{3/2} |n>
\ee
The matrix elements of the standard model will then be modified by the
following form factors:
\be 
<n''| \bar{D}_{ -3t''_{3} -3t''_{0}}^{3/2} \;
D_{-3t'_3- 3t'_0}^{3t} \; D_{-3t_3-3t_0}^{3/2} |n>
\ee
Here  $n$  and  $n''$  take on the values  $0,1,2$  corresponding to the
3 generations in each family of fermions.  These form factors are 2
parameter numerical functions that are in principle observable.  To
calculate them one needs the solitonic factors 
$D^j_{mm^\prime}(a,b,c,d)$ shown in Tables (5.1) and (5.2).  In previous
work we have taken the $|n\rangle$ to be eigenstates of $b$ and $c$
and they have been assumed to be eigenstates of mass.

\begin{table}[h]
\begin{center}
\bf{Table 5.1}
\end{center}
\begin{center}
\begin{tabular}{cccccc}
\underline{$(f_1,f_2,f_3)$} & \underline{$t$} &
\underline{$t_3$} &
\underline{$t_0$} & \underline{$Q$} &
\underline{$D^{3t}_{-3t_{3}-3t_{0}}$} \\
$(e,\mu,\tau)$ & $\frac{1}{2}$ & $-\frac{1}{2}$ & $-\frac{1}{2}$ &
 $-e$ & $D^{3/2}_{\frac{3}{2}\frac{3}{2}} \sim a^{3}$ \\
$(\nu_e,\nu_\mu,\nu_\tau)$ & $\frac{1}{2}$ & $\frac{1}{2}$ &
$-\frac{1}{2}$ & 0  & $D^{3/2}_{-\frac{3}{2}\frac{3}{2}} \sim c^{3}$ \\
$(d,s,b)$ & $\frac{1}{2}$ & $-\frac{1}{2}$ & $\frac{1}{6}$ &
$-\frac{1}{3}e$ &  $D^{3/2}_{\frac{3}{2}-\frac{1}{2}}
\sim a b^{2} $ \\
$(u,c,t)$ & $\frac{1}{2}$ & $\frac{1}{2}$ & $\frac{1}{6}$ &
$\frac{2}{3}e$ &  $D^{3/2}_{-\frac{3}{2} -\frac{1}{2}}
\sim c d^{2}$ \\
\end{tabular}
\end{center}
\end{table}

\begin{table}[h]
\begin{center}
\bf{Table 5.2}
\end{center}
\begin{center}
\begin{tabular}{cccccc}
    &\underline{$t$} & \underline{$t_3$} & \underline{$t_0$} & 
    \underline{$Q$} &            \underline{$D_{-3t_3 -3t_0}^{3t}$}\\
$W^+$  & $1$   &  $1$ & $0$ &$e$  & $D_{-30}^3 \sim c^3 d^3 $ \\
$W^-$  & $1$   & $-1$  & $0$ & $-e$ & $D_{30}^3 \sim a^3 b^3 $ \\
$W^3$  & $1$   & $ 0$ & $0$ & $0$ & $D_{00}^3 \sim f_{3}(b,c)$ \\
$W^0$  & $0$   & $ 0$  & $ 0$ & $0$ &  $D_{00}^0 \sim f_{0}(b,c)$
\end{tabular}
\end{center}
\end{table}
The solitonic factors have been computed according to (3.1) 
and are all
monomials except for the neutral  $W^0$  and  $W^3$.  The numerical
factors  ${\cal A}_{mm'}^j$  have been dropped but may be computed according to
\be
{\cal A}_{mm'}^{j}=  
\left[ \frac{\left< n_+^\prime \right>_{q_1}! \; \left<n_-^\prime\right>_{q_1}!}{\left<n_+\right>_{q_1}!\; \left<n_-\right>_{q_1}!}
\right] ^{\frac{1}{2}} \; 
\left< {\scriptstyle n_+ \atop s} \right>_{q_{1}}
\left< {\scriptstyle n_- \atop t} \right>_{q_{1}} 
\ee
where
\be
\left< {\scriptstyle n \atop s} \right> _q
=
\frac{\left< n\right> _q !}{\left< n-s\right >_q !
\left< s\right >_q ! }  
\; \; \; \mbox{with}\; \;
\langle n\rangle _q = \frac{q^n -1}{q-1}; \;\;\;\;  q_1 =q^{-1}
\ee

Since we require that the fermion-boson interaction be expressed by
(5.1), and that the total action be both $SU(2)\times U(1)$ and
$U_a(1)\times U_b(1)$ invariant, (5.1) and (5.5) must share this
invariance.  Then in view of these invariances and
since $(4.13m)$ and $(4.13m^\prime)$ hold for
${\cal{F}}$, they must also hold for ${\cal{B}}$.  Hence these
relations are not simply conjectured extensions but they are an
essential requirement of the 
electroweak sector of the knot model.  Eq. $(4.13j)$ is not required
but is allowed.

\vskip.5cm

\section{The Dynamics of the Quantum Knot}

$D^j_{mm^\prime}(q(a,b,c,d))$ is a kinematic factor describing a
generic quantum knot.  The corresponding classical knot $(N,w,r)$
has arbitrary size and shape.  To describe the oscillations of this
quantum knot in a field theory one replaces the classical Fourier
modes by quantum oscillators determined by a quantum Hamiltonian.  The
kinematics of the model is determined by $SL_q(2)$ and we shall
restrict the Hamiltonian by the same symmetry.  Then the normal modes 
of the field operators that describe the physical particles are
determined both dynamically and kinematically by $SL_q(2)$.  (It is
similarly possible to restrict both the dynamic and kinematic
dependence of states of the quantized hydrogen atom by a single
(rotation) group (without explicitly introducing the Coulomb 
potential.)$^5$
There is, however, 
no uniquely defined way of dynamically implementing this
symmetry.

If the knot oscillates like the standard quantum mechanical
harmonic oscillator, the Hamiltonian is of the following form:
\be
H = \frac{\hbar\omega}{2}(a\bar a + \bar aa)
\ee
where $\bar a$ and $a$ are raising and lowering operators and
\be
[a,\bar a] = 1
\ee
Since the raising and lowering operators of the $SL_q(2)$ algebra,
that correspond to $\bar a$ and $a$, are $d$ and $a$ respectively,
the analogue of (6.1) in the knot algebra is
\be
H = \frac{\hbar\omega}{2}(ad+da)
\ee
but by the algebra $(A)$
\be
[a,d] = (q-q_1)bc
\ee
and
\be
\frac{1}{2} (ad+da) = 1 + \frac{1}{2}(q+q_1)bc
\ee
We may generalize the $SL_q(2)$ analogue of the 
Hamiltonian of the harmonic oscillator if
we replace (6.3) by a more general function of $ad+da$, 
or of $bc$ by (6.5), or with a still different $H$ by
\be
H = H(b,c) \frac{\hbar\omega}{2}
\ee
Since $b$ and $c$ commute, they have common eigenstates.  
Let $|0\rangle$ be the ground state and let
\begin{eqnarray}
b|0\rangle &=& \beta|0\rangle \\
c|0\rangle &=& \gamma|0\rangle \\
|n\rangle &\sim& d^n|0\rangle
\end{eqnarray}
Then by the algebra $(A)$
\be
b|n\rangle = q^n\beta|n\rangle ~~~\mbox{and} ~~~
c|n\rangle = q^n\gamma|n\rangle
\ee

Let the Hamiltonian of the quantum knot be $H(b,c)$.  Let us consider
the states of this knot defined by $D^j_{mm^\prime}|n\rangle$.  We
may then compute
\begin{eqnarray}
H(b,c)D^j_{mm^\prime}|n\rangle &=& H(b,c)
\left[\sum_{s,t} {\cal{A}}^j_{mm^\prime}\delta(s+t,n^\prime_+)
a^sb^{n_+-s}c^td^{n_--t}\right] |n\rangle \\
&=& D^j_{mm^\prime}H(q_1^{n_a-n_d}b,q_1^{n_a-n_d}c)|n\rangle
\end{eqnarray}
where $n_a$ and $n_d$ are the exponents of $a$ and $d$ respectively,
and $n_a-n_d$ is the same for every term of $D^j_{mm^\prime}$.
Then one has
\be
\begin{array}{rcl}
H(b,c) D^j_{mm^\prime}|n\rangle &=& D^j_{mm^\prime}
H(q_1^{n_a-n_d}q^n\beta,q_1^{n_a-n_d}q^n\gamma)|n\rangle \\
&=& E^j_{mm^\prime}(n){\cal{D}}^j_{mm^\prime}|n\rangle
\end{array}
\ee
where the eigenvalues of $H$ are
\be
E^j_{mm^\prime}(n) = H(\lambda\beta,\lambda\gamma)
\ee
and
\be
\lambda = q^{n-(m+m^\prime)}
\ee
The eigenstates of $H$ are the $D^j_{mm^\prime}|n\rangle$ and the
indices on $D^j_{mm^\prime}$ are the eigenvalues of the integrals of motion.  Since $(m,m^\prime) = 3(-t_3,-t_0)$ by (4.10), we have
by (6.15)
\be
\lambda = \lambda(n,Q_e)
\ee
where
\be
\lambda(n,Q_e) = q^{n+\frac{3}{e} Q_e}
\ee
and where $Q_e$ is the electric charge of the knot.  Here 
$\lambda(n,Q_e)$ is different for each value of $Q_e$, which labels
the fermion family, and each value of $n$, which labels the generation.
For example, $Q_e = -e$ for leptons and $n=0,1,2$, where the $(0,1,2)$
states are electron, muon, and tauon states.  The index $n$ labels
states of different mass, while the operators $a$ and $d$ have matrix
elements connecting adjacent states of different mass and the same
charge.

Finally, in order that the $H$ introduced in (6.6) qualify as the
Hamiltonian of an elementary fermionic knot we shall require that it
be compatible with the fermion mass term in the standard theory,
namely
\be
{\cal{M}} = \bar L\varphi R + \bar R\varphi L
\ee
where $L$ and $R$ are left- and right-chiral Lorentz spinors and
$\varphi$ is the Higgs field, a Lorentz scalar, so that product
$\bar L\varphi R$ is Lorentz invariant.  In the Lagrangian
of the standard model
$L$ and $\varphi$ are isotopic doublets.  $(\bar L\varphi)$ and $R$
are separately isotopic singlets and ${\cal{M}}$ 
is invariant under the
gauged $SU(2)\times U(1)$ group.

In the knot model $L$ is aditionally a fermionic knot with the charge
structure $D^{3/2}_{-3t_3-3t_0}$.  If a knot singlet is assigned to
$\varphi$, then $\varphi$ is neutral (unitary gauge) while the right
chiral spinor must have the same knot state as the left chiral
spinor, namely, $D^{N/2}_{-3t_3-3t_0}$, in order to preserve the
$U_a(1)\times U_b(1)$ invariance.  Then the standard Higgs mechanism
is still possible with $\varphi\sim D^0_{00}$.  

If $L$ and $R$ are now assigned the same internal state, and we treat
the mass term in the same way as the other terms of the Lagrangian,
then we have
\begin{eqnarray}
L &\to& \chi_L(t_3,t_0,n) D^{3/2}_{-3t_3-3t_0}|n\rangle \\
R &\to& \chi_R(t_3,t_0,n) D^{3/2}_{-3t_3-3t_0}|n\rangle
\end{eqnarray}
where $\chi_L(t_3,t_0,n)$ and $\chi_R(t_3,t_0,n)$ are the standard fermionic
chiral fields for the particle labelled $(t_3,t_0,n)$.

Then
\be
{\cal{M}}(t_3,t_0,n) = \langle n|\bar D^{3/2}_{-3t_3-3t_0}
D^{3/2}_{-3t_3-3t_0}|n\rangle (\bar\chi_L\varphi\chi_R +
\bar\chi_R\bar\varphi\chi_L)
\ee
By the argument of the standard theory
\be
\bar\chi_L\varphi\chi_R + \bar\chi_R\bar\varphi\chi_L
\ee
may be reduced to
\be
\rho(\bar\chi_L\chi_R + \bar\chi_R\chi_L) = \rho\bar\chi\chi
\ee
where $\rho$ is the vacuum expectation value of $\varphi$, the Higgs
field.  Then by (6.18)
\be
{\cal{M}}(t_3,t_0,n) = m(t_3,t_0,n)\bar\chi\chi
\ee
and by (6.21)
\be
m(t_3,t_0,n) = \rho(t_3,t_0)
\langle n|\bar D^{3/2}_{-3t_3-3t_0}D^{3/2}_{-3t_3-3t_0}|n\rangle
\ee
which is
compatible with (6.6) and (6.17) as shown, for example, in
Ref. 4, where the energy levels 
given by (6.25) are fixed as polynomial functions 
of $q^n$.

\vskip.5cm

\section{Flavor States}

The states $|n\rangle$ appearing in the form factor (5.6) are
to be interpreted as mass states since they are states of
the general Hamiltonian (6.6).  However, instead of taking the operator
in (5.6) between the states $|n\rangle$, eigenstates of $b$ and $c$,
one may take the same operator between eigenstates of $d$ and $a$ as
follows:
\be
\langle d^\prime|M|a^\prime\rangle = \sum
\langle d^\prime|b^{\prime\prime}\rangle
\langle b^{\prime\prime}|M|b^\prime\rangle
\langle b^\prime|a^\prime\rangle
\ee
where $a^\prime$ and $d^\prime$ are eigenvalues of $a$ and $d$ and
\begin{eqnarray}
|a^\prime\rangle &=& \sum |b^\prime\rangle\langle b^\prime|
a^\prime\rangle \\
|d^\prime\rangle &=& \sum |b^\prime\rangle\langle b^\prime|d^\prime\rangle
\end{eqnarray}
In Ref. (4) with $M$ given by (5.6)
the matrix $\langle n^{\prime\prime}|M|n^\prime\rangle =
\langle b^{\prime\prime}|M|b^\prime\rangle$ describing quark-quark
interactions was computed.  We now 
describe this matrix in the $(a,d)$ instead of the $(b,c)$
representation.  Since the $(d,a)$ operators are emission and
absorption operators while the $(b,c)$ operators define mass states,
we shall describe the $|d^\prime\rangle$ and $|a^\prime\rangle$
states as flavor states. By (7.2) and (7.3) they are superpositions
of mass states. 
Since the eigenstates of the emission and
absorption operators correspond to the coherent
states of the Maxwell field, we shall also refer to 
the corresponding $SL_q(2)$ states as coherent states.

In the photon case the raising operator increases the number of
photons by one.  Here the raising operator increases the number of
$|n+1\rangle$ particles and simultaneously 
decreases the number of $|n\rangle$
particles, in each case also by one.  The lowering operator, as usual,
acts oppositely to the raising operator.  In the photon case the
index $n$ is the number of photons in one mode, while here $n$ refers
to the generation.

\vskip.5cm

\section{The Coherent States of $SU_q(2)$}

We consider the unitary algebra $SU_q(2)$ obtained from $SL_q(2)$ by
setting
\begin{eqnarray}
d &=& \bar a\\
c &=& -q_1\bar b
\end{eqnarray}
Then the $SU_q(2)$ algebra is
\be
\begin{array}{ll}
ab = qba & a\bar a + b\bar b = 1 \\
a\bar b = q\bar ba & \bar aa + q_1^2\bar bb = 1
\end{array}
\ee
From the algebra one has
\be
b\bar a^n = q^n\bar a^nb
\ee
Let $|0\rangle$ be the lowest eigenstate of $b$ and let $\beta$ be the
lowest eigenvalue.  Then
\be
b\cdot\bar a^n|0\rangle = \beta q^n\cdot\bar a^n|0\rangle
\ee
Then the
$\bar a^n|0\rangle$ are eigenstates of $b$ with eigenvalues
$\beta q^n$.  Let
\be
|n\rangle = \bar a^n|0\rangle
\ee
up to a normalization.  Then
\be
\bar a|n\rangle = \bar\lambda_n|n+1\rangle 
\ee
$$
\langle n|a = \langle n+1|\lambda_n \eqno(8.7)^\prime
$$
and
\be
\langle n|a\bar a|n\rangle = |\lambda_n|^2\langle n+1|n+1\rangle
\ee
Normalizing all states by setting $\langle n|n\rangle = 1$, we have
\be
|\lambda_n|^2 = \langle n|a\bar a|n\rangle
\ee
By (8.3)
\be
\begin{array}{rcl}
|\lambda_n|^2 &=& \langle n|1-b\bar b|n\rangle \\
&=& 1-q^{2n}|\beta|^2
\end{array}
\ee
We have assumed that the mass states $|n\rangle$ are eigenstates of
the Hamiltonian:
\be
H|n\rangle = E_n|n\rangle
\ee
where $H$ is a function of
\be
\frac{1}{2} (a\bar a + \bar aa) = 1 - \frac{1}{2}
(1+q_1^2)\bar bb
\ee
The eigenstates of the Hamiltonian are then also eigenstates of 
$\bar bb$.  They are orthogonal since $H$ and $\bar bb$ are
hermitian.

We define the coherent states $|\alpha\rangle$ as eigenstates of the
absorption and emission operators:
\be
a|\alpha\rangle = \alpha|\alpha\rangle 
\ee
\be
\langle\alpha|\bar a = \langle\alpha|\alpha^* 
\ee
To express the coherent states as a superposition of the mass states,
i.e. as
\be
|\alpha\rangle = \sum_n |n\rangle\langle n|\alpha\rangle
\ee
we need the coefficients $\langle n|\alpha\rangle$.  

By (8.7)$^\prime$
\be
\langle n|a|\alpha\rangle = \lambda_n\langle n+1|\alpha\rangle
\ee
and by (8.13)
\be
\langle n|a|\alpha\rangle = \alpha\langle n|\alpha\rangle
\ee
Then by (8.16) and (8.17)
\begin{eqnarray}
\langle n+1|\alpha\rangle &=& \frac{\alpha}{\lambda_n}
\langle n|\alpha\rangle \\
&=& \frac{\alpha}{\lambda_n}\frac{\alpha}{\lambda_{n-1}}
\ldots \langle 0|\alpha\rangle
\end{eqnarray}
where by (8.10)
\be
\lambda_n = |1-q^{2n}|\beta|^2|^{1/2} e^{i\varphi_n}
\ee
Then
\be
\langle n|\alpha\rangle = \frac{\alpha^n}
{{\displaystyle\prod^{n-1}_0}\lambda_s}
\langle 0|\alpha\rangle  \qquad \qquad n\geq 1
\ee
and
\be
\sum_{n\geq 0}\langle \alpha|n\rangle\langle n|\alpha\rangle =
\sum_{n\geq 1} \frac{|\alpha|^{2n}}
{|{\displaystyle\prod^{n-1}_0}\lambda_s|^2}
|\langle 0|\alpha\rangle|^2 + |\langle 0|\alpha\rangle|^2
\ee
or if the $|n\rangle$ are complete
\be
\langle\alpha|\alpha\rangle = \sum_{n\geq 1}
\frac{|\alpha|^{2n}}{{\displaystyle\prod^{n-1}_0} |\lambda_s|^2}
|\langle 0|\alpha\rangle|^2 + |\langle 0|\alpha\rangle|^2
\ee
Then, normalizing $\langle\alpha|\alpha\rangle = 1$, one has
\be
|\langle 0|\alpha\rangle|^{-2} = \sum_{n\geq 1}
\frac{|\alpha|^{2n}}{|{\displaystyle\prod^{n-1}_0}\lambda_s|^2} + 1
\ee
As usual $\langle\alpha|n\rangle$ is the adjoint of $\langle n|\alpha\rangle$, but the bracket $\langle n|\alpha\rangle$ between the mass
and flavor states is not unitary.

\vskip.5cm

\section{The KM and the PMNS Matrices}

Since there are only three generations in each family of fermions,
there are only three mass states, which we label $n=0,1,2$; and there are only three flavor states 
which we label by $\alpha_0,\alpha_1,\alpha_2$.  
Because we identify the
flavor states as the coherent states, we have by (8.20) and (8.21)
\begin{eqnarray}
\langle 1|\alpha_i\rangle &=& \frac{\alpha_i}
{[1-|\beta|^2]^{1/2}} \langle 0|\alpha_i\rangle \qquad i = 0,1,2 \\
\langle 2|\alpha_i\rangle &=& \frac{\alpha_i^2}
{[(1-|\beta|^2)(1-q^2|\beta|^2)]^{1/2}} \langle 0|\alpha_i\rangle
\end{eqnarray}
By (8.15)
\be
|\alpha_i\rangle = |0\rangle\langle 0|\alpha_i\rangle +
|1\rangle\langle 1|\alpha_i\rangle + |2\rangle\langle 2|\alpha_i\rangle
\qquad i = 0,1,2
\ee
We shall take the mass states $|n\rangle$ orthonormal.  Then we have
\be
\langle\alpha_i|\alpha_i\rangle = 
|\langle 0|\alpha_i\rangle|^2 + |\langle 1|\alpha_i\rangle|^2 +
|\langle 2|\alpha_i\rangle|^2
\ee
Normalizing the coherent states, $\langle\alpha_i|\alpha_i\rangle = N_i$,
one has by (9.1), (9.2) and (9.4)
\be
|\langle 0|\alpha_i\rangle| =N_i^{-\frac{1}{2}}
\left[1 + \frac{|\alpha_i^2|}{1-|\beta|^2} + 
\frac{|\alpha_i^4|}{(1-|\beta|^2)(1-q^2|\beta|^2)}\right]^{-1/2}
\ee
In our earlier work the elements of the form factor (5.6) were expressed
in the $|n\rangle$ or mass basis where the $|n\rangle$ are
eigenstates of $b$ and $c$.  We now express the same operators in
the $|\alpha\rangle$ or coherent basis, 
i.e. eigenfunctions of the absorption $(a)$ and
creation operators $(\bar a)$.  The elements of the form factor in this
basis are
\begin{eqnarray}
\langle u(i)|W^+|d(j)\rangle &=& \sum_{nn^\prime}
\langle u(i)|u(n)\rangle\langle u(n)|W^+|d(n^\prime)\rangle
\langle d(n^\prime)|d(j)\rangle \\
\langle d(j)|W^-|u(i)\rangle &=& \sum_{nn^\prime}
\langle d(j)|d(n)\rangle\langle d(n)|W^-|u\langle n^\prime|\rangle
\langle u(n^\prime)|u(i)\rangle
\end{eqnarray}
where $u(i)$ and $d(j)$ are the up $(u,c,t)$ and down $(d,s,b)$ quark
triplets.  Here $i$ and $j$ describe flavor states while $n$ and 
$n^\prime$ describe mass states.
In these equations $W^+$ and $W^-$ refer to charged $W$
fields and are represented by $D^3_{-30}$ and $D^3_{30}$
respectively as in Table 5.2 and Eq. (5.6).

The associated form factors are special cases of (5.6).  Corresponding
to (9.6) and (9.7) we have
\begin{eqnarray}
\langle u(n)|W^+|d(n^\prime)\rangle &\sim& \langle n|
\bar D^{3/2}_{-\frac{3}{2}-\frac{1}{2}}D^3_{-30}D^{3/2}_{\frac{3}{2}
-\frac{1}{2}}|n^\prime\rangle \quad \mbox{and}  \\
\langle d(n)|W^-|u(n^\prime)\rangle &\sim& \langle n|
\bar D^{3/2}_{\frac{3}{2}-\frac{1}{2}} D^3_{30}
D^{3/2}_{-\frac{3}{2}-\frac{1}{2}}|n^\prime\rangle
\end{eqnarray}
With the same model for the PMNS matrix, the form factor is
\be
\langle n|\bar D^{3/2}_{-\frac{3}{2}\frac{3}{2}} D^j_{00}
D^{3/2}_{-\frac{3}{2}\frac{3}{2}}|n^\prime\rangle
\ee
where $n=0,1,2$ label the three generations, e.g. the $e,\mu$ and
$\tau$ neutrino states.  Here the $|n\rangle$ represent mass states.
Since charge and hypercharge are conserved, the product 
$\bar D^{3/2}D^jD^{3/2}$ is neutral.  
It therefore lies in the $(b,\bar b)$
subalgebra and has no off-diagonal elements.  Then
\begin{eqnarray}
\langle n^\prime|\bar D^{3/2}_{m^{\prime\prime}p^{\prime\prime}}
D^j_{m^\prime p^\prime}D^{3/2}_{mp}|n\rangle &=&
\langle n^\prime|F(b,\bar b)|n\rangle \\
&=& \delta(n^\prime,n) F_n
\end{eqnarray}
In terms of flavor states one then has
\be
\langle i|F(b,\bar b)|j\rangle = \sum_n\langle i|n\rangle
F_n\langle n|j\rangle
\ee
The form factors, which are diagonal in mass states
$|n\rangle$, are not diagonal in the flavor states
$|i\rangle$.

In both the KM and PMNS cases one makes use of the matrix
$\langle n|\alpha\rangle$ that describes the mixing of mass states
in the flavor states.  The observational consequences are, however,
very different in the two cases: the KM matrix describes transitions
between quarks of different charge that are mediated by the $W^\pm$
field and in this case the $\langle n|\alpha\rangle$ matrix simply
changes the basis from mass to flavor states; the PMNS matrix, on the
other hand, describes neutral
transitions between neutrinos of different mass.
In both the quark and neutrino cases the different mass states travel
at different velocities and oscillate at different frequencies but
only in the neutrino case does the particle move far enough for the
interference to be observed.

The probability of a neutrino being detected after the time $t$ in the
flavor states $\nu_j$ if it is emitted in the flavor state $\nu_i$
is
\be
P_{i\to j} = |\langle\nu_j|\nu_i(t)\rangle|^2
\ee
where the flavor states are superpositions of mass states $\nu_n$:
\be
\nu_i = \sum U_{in}\nu_n
\ee

The mass states are orthonormal
\be
\langle \nu_n,\nu_{n^\prime}\rangle = \delta_{nn^\prime}
\ee
and propagate according to
\be
\nu_n(t) = e^{i(E_nt-\vec p_n\vec x)}\nu_n(0)
\ee
If $p \gg m$
\be
\nu_n(t) \sim e^{-im^2_nt/2E_n}\nu_n(0)
\ee
Then
\be
P_{i\to j}(t) = |U^*_{jn}U_{in}e^{-im_n^2t/2E_n}|^2
\ee
Constructive interference makes it possible for a neutrino created
with a given flavor to change its flavor during propagation.

\vskip.5cm

\section{An Alternative Implementation of the $SL_q(2)$ 
Symmetry$^6$}

The previous sections of this paper have been based on discussions
of the $SL_q(2)$ algebra where the mass states are identified with
eigenstates of $b$ and $c$ and an associated Hamiltonian.  The
eigenvalues of this Hamiltonian are functions of 
$q^n$ rather than $n$; 
the latter measures the eigenvalues of the Hamiltonian of the
standard harmonic oscillator, and the $q^n$ form a geometric rather
than an arithmetical progression.  We shall now describe an
implementation of the $SL_q(2)$ symmetry that is closer to the
standard procedures which are dependent on the familiar quantum
oscillator.  In this way one arrives at a different presentation of
the flavor states, but the energy levels still turn out to depend on
$q^n$.

The invariant matrix of $SL_q(2)$ is a 2-dimensional square root of
-1, namely
\be
\epsilon_q = \left(\matrix{0 & q^{-1/2} \cr
-q^{1/2} & 0 \cr} \right) 
\ee
Any 2-dimensional representation of $SL_q(2)$:
\be
T = \left(\matrix{a & b \cr c & d \cr} \right)
\ee
satisfies
\be
T\epsilon_q T^t = T^t\epsilon_qT = \epsilon_q
\ee
and the elements of $T$ satisfy the knot algebra $(A)$.

One may define a 2-dimensional vector basis $\Lambda$ of $SL_q(2)$ by
\be
\Lambda^t\epsilon_q\Lambda = q^{-1/2}
\ee
Then (10.4) is invariant under $\Lambda^\prime = T\Lambda$.
Let
\be
\Lambda = \left(\matrix{D_x \cr x \cr} \right)
\ee
Then
\be
D_xx - qxD_x = 1
\ee
This equation will be satisfied if $D_x$ is chosen as a $q$
difference operator:
\be
D_x\psi(x) = \frac{\psi(qx)-\psi(x)}{qx-x}
\ee
Then $xD_x$ is a ``basic dilatation operator":
\begin{eqnarray}
xD_x &=& \left\langle x\frac{d}{dx}\right\rangle \\
&=& \frac{q^{x\frac{d}{dx}}-1}{q-1}
\end{eqnarray}
If we introduce
\be
P_x = \frac{\hbar}{i} D_x
\ee
then
\be
(P_xx-qxP_x)\psi(x) = \frac{\hbar}{i} \psi(x)
\ee
If $q=1$, (10.11) is the Heisenberg commutator.

We also introduce the $q$-commutator and rewrite (10.6) as
\be
[P_x,x]_q = -i\hbar
\ee
We may quantize by (10.4) with $\Lambda$ given by (10.5) and (10.10).We may also quantize by any $\Lambda^\prime$ related to $\Lambda$
by
\be
\Lambda^\prime = T\Lambda \qquad T\epsilon SL_q(2)
\ee
If $q$ is near unity (as it must be insofar as the standard theory
$(q=1)$ is approximately correct) then $q=1+\epsilon$ and by 10.7)
\be
D_x\psi(x) = \frac{\psi(x+\epsilon x)-\psi(x)}{\epsilon x}
\ee
and $D_x$ resembles the difference operator on a lattice space,
and $q$ may play the role of a dimensionless regulator.

Let us next apply this method of quantization to a harmonic oscillator.

\vskip.5cm

\section{The $q$-Quantized Oscillator$^6$}

Let us quantize according to (10.4) and (10.13) by taking
\be
\Lambda^\prime = Z = \left(\matrix{z \cr \bar z \cr}\right)
\ee
Then
\be
Z^t\epsilon_q Z = q^{-1/2}
\ee
or
\be
z\bar z-q\bar zz = 1
\ee
or
\be
[z,\bar z]_q = 1
\ee
Let us interpret $z$ and $\bar z$ as absorption and emission
operators and retain the usual harmonic oscillator Hamiltonian written
in terms of these operators:
\be
H = \frac{\hbar\omega}{2} (z\bar z+\bar zz)
\ee
Then if we modify the Heisenberg equation of motion in accordance
with (11.4) we have
\be
i\hbar \dot z = [z,H]_q
\ee
Assuming (11.5) and (11.6) we find the usual harmonic dependence
\be
z \sim e^{i\omega t}
\ee
Denote the eigenstates of $H$ by $|n\rangle$.  Then $z$ and
$\bar z$ will satisfy (11.3) if
\begin{eqnarray}
z|n\rangle = \langle n\rangle_q^{1/2}|n-1\rangle & &
\langle n|\bar z = \langle n-1|\langle n\rangle_q^{1/2} \\
\bar z|n\rangle = \langle n+1\rangle_q^{1/2}|n+1\rangle & &
\langle n|z = \langle n+1\rangle_q^{1/2}\langle n+1|
\end{eqnarray}
where $\langle n\rangle_q$ is the ``basic number"
\be
\langle n\rangle_q = \frac{q^n-1}{q-1}
\ee
Then
\begin{eqnarray}
z\bar z|n\rangle &=& \langle n+1\rangle_q|n\rangle \\
\bar zz|n\rangle &=& \langle n\rangle_q|n\rangle
\end{eqnarray}
and
\be
(z\bar z-q\bar zz)|n\rangle = (\langle n+1\rangle_q-
q\langle n\rangle_q)|n\rangle = |n\rangle
\ee
Therefore the commutator (11.3) is satisfied.  By (11.5)
\be
H|n\rangle = \frac{\hbar\omega}{2}(\langle n+1\rangle_q + \langle n\rangle_q)|n\rangle
\ee

The eigenvalues of $H$ in the limit $q=1$ are
\[
\frac{1}{2} \hbar\omega(2n+1)
\]
in agreement with the standard harmonic oscillator.
For general values of $q$, however, one has
\[
\frac{1}{2}\left[\langle n+1\rangle_q+\langle n\rangle_q\right]
= \frac{1}{2} \frac{q^{n+1}+q^n-2}{q-1}
\]
in agreement with (6.17) in its dependence on $q^n$ rather than $n$.
In fact, one may define on the algebra (8.3) 
as a special case of (6.6), a linear function of
$b$, namely
\[
H_b = \left(\frac{1}{2} \frac{q+1}{q-1} \frac{b}{\beta} -
\frac{1}{q-1}\right) \hbar\omega
\]
such that
\[
H_b|n\rangle = E_n|n\rangle
\]
where the $|n\rangle$ are the eigenstates of $b$, with eigenvalues
$\beta q^n$ by (8.5), and
\begin{eqnarray*}
E_n &=& \left(\frac{1}{2} \frac{q+1}{q-1} 
q^n - \frac{1}{q-1}\right) \hbar\omega \\
&=& \frac{1}{2}\left(\langle n+1\rangle_q + \langle n\rangle_q\right)
\hbar\omega
\end{eqnarray*}
with the spectra and eigenstates of the Hamiltonian of the
$z$-oscillator, namely
\[
H_z = \frac{1}{2}(\bar zz + z\bar z)\hbar\omega
\]
with
\[
z\bar z-q\bar zz = 1
\]
according to (11.3), (11.5) and (11.14).

The coherent states are eigenstates of the absorption operator.  Denote the coherent states by $|\zeta\rangle$.  Then
\be
z|\zeta\rangle = \zeta|\zeta\rangle
\ee
We express $|\zeta\rangle$ as a superposition of the eigenstates
$|n\rangle$ of $H$ i.e.,
\be
|\zeta\rangle = \sum |n\rangle\langle n|\zeta\rangle
\ee
The $|n\rangle$ and $|\zeta\rangle$ states are again states of mass
and flavor respectively.

To compute $\langle n|\zeta\rangle$ note
\be
\langle n|z|\zeta\rangle =
\langle n+1\rangle_q^{1/2}\langle n+1|\zeta\rangle
\ee
and
\be
\langle n|z|\zeta\rangle = \zeta\langle n|\zeta\rangle
\ee
Then
\be
\langle  n+1|\zeta\rangle = \frac{\zeta}{\langle n+1\rangle_q^{1/2}}
\langle n|\zeta\rangle
\ee
and
\be
\langle n|\zeta\rangle = \frac{\zeta^n}{\langle n\rangle_q^{1/2}!}
\langle 0|\zeta\rangle
\ee
One may normalize by setting
\be
\langle 0|\zeta\rangle = e_q^{-|\zeta|^2}
\ee
where
\be
e_q^{|\zeta|^2} = \sum^\infty_{n=0}
\frac{|\zeta|^{2n}}{\langle n_q\rangle!}
\ee
is the twisted exponential.  Then
\be
\langle\zeta|\zeta\rangle = \sum\langle\zeta|n\rangle\langle n|\zeta
\rangle = 1
\ee

If $q=1$ then the basic numbers $\langle n\rangle_q$ are replaced
by $n$ and one recovers the familiar results for the coherent
states of the harmonic oscillator.  Since there are again only three
ococupied states, one has
\be
|\zeta_i\rangle = |0\rangle\langle 0|\zeta_i\rangle + |1\rangle
\langle 1|\zeta_i\rangle + |2\rangle\langle 2|\zeta_i\rangle
\ee
Since the mass states are again orthonormal, one has
\be
\langle\zeta_i|\zeta_i\rangle = |\langle 0|\zeta_i\rangle|^2 +
|\langle 1|\zeta_i\rangle|^2 + |\langle 2|\zeta_i\rangle|^2
\ee
where by (11.20)
\begin{eqnarray}
\langle 1|\zeta_i\rangle &=& \zeta_i\langle 0|\zeta_i\rangle \\
\langle 2|\zeta_i\rangle &=& \frac{\zeta_i^2}{(1+q)^{1/2}}
\langle 0|\zeta_i\rangle
\end{eqnarray}

Normalize $\langle \zeta|\zeta\rangle = 1$.  Then by (11.26)-(11.28)
\be
\langle 0|\zeta_i\rangle =
\left[1+|\zeta_i|^2 +
\frac{|\zeta_i|^4}{(1+q)}\right]^{-1/2}
\ee

We now have two representations of the mass-flavor mixing matrix: 
either $\langle n|\alpha\rangle$ in Section 9 or $\langle n|
\zeta\rangle$ in Section 11.  Both representations are allowed by
$SU_q(2)$. 

\vskip.5cm

\section{The Knot Parameterization of the CKM Matrix}

\noindent The three mass states of the model $\left | n \right.\rangle$, n = 0,1,2, representing the three generations, are eigenstates of the hermitian operator $\overline{b}b$.  These states are orthogonal with real eigenvalues.  The three flavor states $\left | i \right. \rangle$, i = 0,1,2, appear as eigenstates of the non-hermitian operator $a$.  The flavor states are then superpositions of the mass states:

\begin{quote}
\begin{center}
$\left | i \right. \rangle = \sum \left | n \right. \rangle\langle n \left | \right. i \hspace{0.5mm}  \rangle$\marginpar{(12.1)}
\end{center}
\end{quote}

\noindent The absolute values of the elements of the mixing matrix, $\left | \langle n \right| i \hspace{0.5mm}\rangle \! \left | \right.$, are the magnitudes of the Cabbibo-Kobayashi-Maskawa matrix elements.  Given the CKM matrix we here describe and partially determine the parameters of the knot model.  Denote the three complex eigenvalues of $a$ by $\alpha_i$, i = 0, 1, 2, and express the length of each flavor state by

\begin{quote}
\begin{center}
$\langle i \left | i \right. \rangle = N_i$ \marginpar{(12.2)}
\end{center}
\end{quote}

\noindent Normalize the length of each mass state by

\begin{quote}
\begin{center}
$\langle n \left | \right. n \rangle = 1$ \marginpar{(12.3)}
\end{center}
\end{quote}

\noindent We shall express the elements of the CKM matrix in terms of $\alpha_i$, $N_i$, and two constants, $q$ and $\beta$, where $q$ fixes the algebra and $\beta$ is the eigenvalue of $b$ on its ground state.

\noindent \hspace{5mm} Since the $\left | \right. \! n  \rangle$ are orthonormal, we have

\begin{quote}
\begin{center}
$N_i = \sum_{n=0}^{2} \langle \hspace{0.5mm} i \left | \right. n \hspace{0.5mm}\rangle \langle \hspace{0.5mm}n \left | \right. i \hspace{0.5mm}\rangle \hspace{10mm}  i = 0, 1, 2$ \marginpar{(12.4)}
\end{center}
\end{quote}

\noindent Then by (8.23) and (12.4)

\begin{quote}
\begin{center}
$N_i = \left | \langle \hspace{0.5mm} 0 \right. \left | \hspace{0.5mm} i \hspace{0.5mm} \rangle \right | ^2 + \left | \langle \hspace{0.5mm} 0 \right | i \hspace{0.5mm}\rangle \left | \right. ^2 \sum_{n=1}^{2} \frac{\left | \alpha_i \right | ^{2n}}{\prod_{0}^{n-1} \left | \lambda_s \right | ^2}  \hspace{10mm} i = 0, 1, 2$ \marginpar{(12.5)}
\end{center}
\end{quote}

\noindent and

\begin{quote}
\begin{center}
$\left | \langle 0 \right | i \rangle \!\left | \right. = N_i^{\frac{1}{2}} \left[ 1 + \left | \frac{\alpha_i}{\lambda_0} \right |^2 +\left | \frac{\alpha_i}{\lambda_0} \right |^4 \left| \frac{\lambda_0}{\lambda_1} \right |^2 \right]^{-\frac{1}{2}} \hspace{10mm} i = 0, 1, 2$ \marginpar{(12.6)}
\end{center}
\end{quote}

\noindent and again by (8.21) and setting $\langle 0 \left | \right. \! \alpha \rangle = \left | \langle 0 \right | \alpha \rangle \! \left | \right.$, we have

\begin{quote}
\begin{center}
$\langle n \left| \right. i \rangle = N_i^{\frac{1}{2}} g(\left |x_i\right|)\frac{\alpha_i^n}{\prod_0^{n-1} \lambda_s} \hspace{10mm} n = 1, 2 \hspace{5mm} i = 0, 1, 2$ \marginpar{(12.7a)}
\end{center}
\end{quote}

\noindent where

\begin{quote}
\begin{center}
$g(x) = \left[ 1 + x^2 + \left | \frac{\lambda_0}{\lambda_1}\right|^2 x^4 \right]^{-\frac{1}{2}}$  \marginpar{(12.7b)}
\end{center}
\end{quote}

\noindent with

\begin{quote}
\begin{center}
$x_i = \frac{\alpha_i}{\lambda_0}$ \marginpar{(12.7c)}
\end{center}
\end{quote}

\noindent If we do not set $\langle 0 \left | \right. \! \alpha_i \rangle = \left | \langle 0 \right | \alpha_i  \rangle \! \left | \right.$, there are three additional phase factors in (12.7).

By (12.7) the mixing matrix $\langle n \left | i \right. \rangle $ is expressed in terms of

\begin{quote}
(a) the eigenvalues, $\alpha_i$, of the absorption-emission operator, a,
\end{quote}
\begin{quote}
(b) the norms, $N_i$, of the eigenstates (the flavor states) of these operators,
\end{quote}
\begin{quote}
(c) the matrix elements $(\langle n \left | a \right | n+1 \rangle = \lambda_n)$ of these same operators, a, between neighboring mass states.
\end{quote}

The $\lambda_n$ in turn is given by (8.10) as a function of $q$ and $\beta$, the knot parameters.

The mixing matrix $\langle n \left | \right. \! i \rangle$, as given by (12.7), is shown in Table 1.

\begin{quote}
\begin{center}
\underline{Table 1}

Elements of the $\langle n \left | \right. \! i \rangle$ Matrix

\begin{tabular}{|c|c|c|c|}
\hline
$n \setminus i$  & 0 & 1 & 2 \\
\hline
0 & $N_0^{\frac{1}{2}}g\left(\left|\frac{\alpha_0}{\lambda_0} \right |\right)$ & $N_1^{\frac{1}{2}}g\left(\left|\frac{\alpha_1}{\lambda_0} \right |\right)$ & $N_2^{\frac{1}{2}}g\left(\left|\frac{\alpha_2}{\lambda_0} \right |\right)$\\
\hline
1 & $N_0^{\frac{1}{2}}\frac{\alpha_0}{\lambda_0}g\left(\left|\frac{\alpha_0}{\lambda_0} \right |\right)$ & $N_1^{\frac{1}{2}}\frac{\alpha_1}{\lambda_0}g\left(\left|\frac{\alpha_1}{\lambda_0} \right |\right)$ & $N_2^{\frac{1}{2}}\frac{\alpha_2}{\lambda_0}g\left(\left|\frac{\alpha_2}{\lambda_0} \right |\right)$ \\
\hline
2 & $N_0^{\frac{1}{2}}\left(\frac{\alpha_0}{\lambda_0}\right)^2\left(\frac{\lambda_0}{\lambda_1} \right)_0g\left(\left|\frac{\alpha_0}{\lambda_0} \right |\right)$ & $N_1^{\frac{1}{2}}\left(\frac{\alpha_1}{\lambda_0}\right)^2\left(\frac{\lambda_0}{\lambda_1} \right)_1g\left(\left|\frac{\alpha_1}{\lambda_0} \right |\right)$ & $N_2^{\frac{1}{2}}\left(\frac{\alpha_2}{\lambda_0}\right)^2\left(\frac{\lambda_0}{\lambda_1} \right)_2g\left(\left|\frac{\alpha_2}{\lambda_0} \right |\right)$\\
\hline
\end{tabular}
\end{center}
\end{quote}
 
\vspace{10mm}
\noindent \hspace{5mm}
The absolute values $\left | \langle n \right| i \rangle \! \left|\right.$ must agree with the magnitudes of the elements of the CKM matrix in Table 2.

\begin{quote}
\begin{center}
\underline{Table 2}
\\
\vspace{5mm}
The CKM Matrix
\vspace{2mm}

\begin{tabular}{|c|c|c|c|}
\hline
n $\setminus$ i & 0 & 1 & 2\\
\hline
0 & 0.97428 & 0.2253 & 0.00347\\
\hline
1 & 0.2252 & 0.97345 & 0.0410\\
\hline
2 & 0.00862 & 0.0403 & 0.999\\
\hline

\end{tabular}
\end{center}
\end{quote}
\vspace{5mm}

Equating the matrix elements $\langle n \left | \right. i \rangle$ of Table 1 to the CKM matrix elements of Table 2, one finds the numerical values of $\left | \frac{\alpha_i}{ \lambda_0} \right | $ and $\left | \frac{\lambda_0}{\lambda_i} \right |_i$ as follows:

\begin{quote}
\begin{center}
$\left | \frac{\alpha_i}{\lambda_i} \right | = \frac{M_{1i}}{M_{0i}}$
\end{center}
\end{quote}

\noindent and

\begin{quote}
\begin{center}
$\left | \frac{\lambda_o}{\lambda_1} \right |_i \hspace{2mm} \left | \frac{\alpha_i}{\lambda_0} \right | = \frac{M_{2i}}{M_{1i}}$
\end{center}
\end{quote}

\noindent where the $M_{ni}$ are the elements of the CKM matrix.  After the numerical values of $\left | \frac{alpha_i}{\lambda_0} \right |$ and $\left | \frac{lambda_0}{\lambda_i} \right |_i$ are found, the numerical values of $N_0$, $N_1$, and $N_2$ are determined by requiring a complete match between the matrix elements in Tables 1 and 2:

\begin{quote}
\begin{center}
$\left | \langle n \right | i  \rangle \left | \! \right. = M_{ni}^{CKM}$
\end{center}
\end{quote}

\hspace{5mm} One then finds for the eigenvalues and normalizations of the flavor states the values displayed in Table 3.

\begin{quote}
\begin{center}
\underline{Table 3}

\begin{tabular}{cccc}
 $\left| \right.$eigenvalues $\!\left|\right.$: & $\left| \frac{\alpha_0}{\lambda_0} \right| = 0.231$  $\left| \frac{\alpha_1}{\lambda_0} \right| = 4.32 $  $ \left| \frac{\alpha_2}{\lambda_0} \right| = 11.8 $ \\

norms : & $N_0^{\frac{1}{2}} = (0.998, 0.997, 0.997) \cong 1$ \\ & $ N_1^{\frac{1}{2}} = (1.07,1.08,1.08) \cong 1$& \\ & $N_2^{\frac{1}{2}} = (0.697, 0.698, 0.696) \cong 0.7$

\end{tabular}
\end{center}
\end{quote}

\noindent There is close agreement among the three values of the three $N_i ( = \langle i \left | \right. i \rangle$) as determined by the three rows of the CKM matrix.

\noindent \hspace{5mm}
The ratio $\left| \frac{\lambda_0}{\lambda_1} \right| = \left( \frac{1-\beta^2}{1-q^2\beta^2} \right)^{\frac{1}{2}}$ appears as a factor in only the $\langle 2 \left|\right. \! 0 \hspace{0.5mm}\rangle$, $\langle 2 \left|\right. \! 1 \hspace{0.5mm} \rangle$, and $\langle 2 \left|\right. \! 2 \hspace{0.5mm} \rangle$ elements of $ \langle n \left | \right. i \rangle$ and is shown in Table 4.

\begin{quote}
\begin{center}
\underline{Table 4}
\end{center}
\begin{center}
\begin{tabular}{cccc}

If $\langle 2 \left| \right.\!  i \hspace{0.5mm}\rangle =$ \hspace{2mm}& $\langle 2 \left|\right. \! 0 \hspace{0.5mm} \rangle$ \hspace{2mm}& $\langle 2 \left|\right.\! 1 \hspace{0.5mm}\! \rangle $ \hspace{2mm}& $ \langle 2 \left| \right. \! 2 \hspace{0.5mm} \rangle$\\
then $\left| \frac{\lambda_0}{\lambda_1} \right|_i = \hspace{2mm}$ & 0.166 \hspace{2mm} & 0.00958 \hspace{2mm} & 2.07 \hspace{2mm}\\

\end{tabular}
\end{center}
\end{quote}
\vspace{4mm}
\noindent It is possible to approximately describe the CKM matrix with only four parameters, as one sees, for example, in the Wolfenstein parameterization.  Since there are in the knot description more than four parameters, these must be approximately related.  The larger number of free parameters should also describe a hypothetical preon substructure.  We note that the eigenvalues of the flavor operators (emission - absorption operators) are complex, and that $\lambda$ and the eigenvalues of $b$, may also be taken as complex, so that the usual signal for the violation of T is present.

\noindent \hspace{5mm}
In Table 4, one finds

\begin{quote}
\begin{center}
$R \equiv \left| \frac{\lambda_0}{\lambda_1} \right| ^2 = \frac{1-\beta^2}{1-q^2\beta^2}$ \marginpar{(12.8)}

\end{center}
\end{quote}

\noindent One may rewrite (12.8) as

\begin{quote}
\begin{center}

$\beta^2 = \frac{1-R}{1-Rq^2} > 0$ \marginpar{(12.9)}

\end{center}
\end{quote}

\noindent or as

\begin{quote}
\begin{center}

$q^2 = \frac{R-1+\beta^2}{R\beta^2} > 0$ \marginpar{(12.10)}

\end{center}
\end{quote}

\noindent Since $\beta$ and $q$ are real, Eqn. (12.9) and (12.10) put limits on $q^2$ and $\beta^2$ respectively.  For example, if $ 0 < R < 1$, then $q^2 < 1/R$ and $\beta^2 > 1-R$.  Note that one solution of (12.8) for all values of R is

\begin{quote}
\begin{center}
$\beta = q = 1$ \marginpar{(12.11)}
\end{center}
\end{quote}

\noindent There is a continuum of solutions in the neighborhood of 

\begin{quote}
\begin{center}
$(\beta, q) = (1,1)$ \marginpar{(12.12)}
\end{center}
\end{quote}

\noindent For example, let

\begin{quote}
\begin{center}
$\beta^2 = 1 - \epsilon \hspace{10mm} 0 < \epsilon \ll1$ \marginpar{(12.13)}
\end{center}
\end{quote}

\noindent then

\begin{quote}
\begin{center}
$q \cong 1 + \left( 1 - \frac{1}{R}\right) \epsilon$ \marginpar{(12.14)}
\end{center}
\end{quote}

\noindent where $R$ is taken from Table 4.

The knot model is based on the successful characterization of the twelve elementary fermions as three states of excitation of the four quantum knots representing leptons, neutrinos, up quarks, and down quarks.  This model introduces form factors that multiply the matrix elements for interactions between the fermions.  These form factors have been computed in previous work$^{(4)}$ and are compatible with experiment if $q \cong 1$.  With the same model we have in this paper attributed the Cabibbo mixing to the knot degrees of freedom of the elementary fermions.  By comparing with the CKM matrix one again finds $q \cong 1$, as well as weak limitations on the parameters describing the knot model.

\section{References}

\begin{enumerate}
\item N. Cabbibo, Phys. Rev. Lett. {\bf 10}, 531 (1963);
M. Kobayashi and K. Maskawa, Prog. Theor. Phys. {\bf 49}, 282 (1972);
B. Pontecorvo, JETP {\bf 6}, 429 (1957);
Z. Maki, M. Nakagawa, and S. Sakata, Prog. Theor. Phys. {\bf 28},
870 (1962).
\item R. J. Finkelstein, Int. J. Mod. Phys. A{\bf 24}, 2307 (2009).
\item R. J. Finkelstein, ArXiv: hep.th/0701124 v1.
\item A. C. Cadavid and R. J. Finkelstein, {\it ibid.} A{\bf 21},
4269 (2006).
\item R. J. Finkelstein, J. Math. Phys. {\bf 8}, 443 (1967).
\item R. J. Finkelstein and Eran Marcus, J. Math. Phys. {\bf 36},
2652 (1995).
\end{enumerate}

\end{document}